\documentclass[11pt]{amsart}
\usepackage{geometry}                
\usepackage{amsmath}
\usepackage{amssymb}
\usepackage{epstopdf}
\usepackage{color,xcolor}
\usepackage{hyperref}
\usepackage{url}
\usepackage{graphicx}
\newcommand{\beq}{\begin{equation}}
\newcommand{\eeq}{\end{equation}}

\title{Integral invariants in flat superspace}
\author{Michael Movshev, Albert Schwarz, Renjun Xu}

\begin{document}
\begin{abstract} {We are solving for the case of flat superspace some homological problems that were formulated by Berkovits and Howe. (Our considerations can be applied also to the case of supertorus.) These problems arise in the attempt to construct  integrals  invariant with respect to supersymmetry. They appear also in other situations, in particular, in the pure spinor formalism in supergravity.}
\end{abstract}
\maketitle
\section{Introduction}

In present paper we are solving for the case of flat superspace some homological problems that were formulated in the paper \cite {BH}. (Our considerations can be applied also to the case of supertorus.) These problems arise in the attempt to construct  integrals  invariant with respect to supersymmetry and in other situations.

Let us  consider a flat superspace with coordinates $z^M \sim (x^m, \theta^{\alpha})$.  The supersymmetry Lie algebra $susy$ is generated by transformations
$$[e_{\alpha},e_{\beta}]_+=\gamma^m_{\alpha\beta}P_m, \text{ with differential }d=\gamma^m_{\alpha\beta}t^\alpha t^\beta\frac{\partial}{\partial C^m}$$
Here $e_{\alpha}$ are generators acting on the space of $(x^m, \theta^{\alpha})$, $P_m$ is the translation operator, and $\gamma^m_{\alpha\beta}$ are Dirac Gamma matrices.

To construct integrals invariant with respect to supersymmetry  one should find closed differential forms expressed in terms of physical fields.
This is the homological problem we are trying to solve (see \cite {BH} for more details).

We consider a basis $E^m=E^m_M dz^M,E^{\alpha}=E^\alpha_M dz^M $ in the space of one-forms given by formulas $E^\alpha=d\theta^\alpha,$ $E^m=dx^m+\Gamma^m_{\alpha\beta}\theta^\alpha d\theta^\beta.$
Every differential form has a unique decomposition
$\omega=\sum\omega_{p,q}$, where
$$\omega_{p,q}=\frac{1}{p!q!}E^{\beta_q}\cdots E^{\beta_1}E^{a_p}\cdots E^{a_1}\omega_{a_1\cdots a_p \beta_1 \cdots \beta_q}(x,\theta)$$
In other words, the space $\Omega$ of all forms is a direct sum of subspaces $\Omega_{p,q}.$

We can consider differential forms as functions depending on even variables $x^m, E^{\alpha}$ and odd variables $\theta ^{\alpha}, E^m$. Using the relations $dE^{\alpha}=0, dE^m=  \Gamma^m_{\alpha\beta} E^{\alpha}E^{\beta}$ we obtain a representation of
the exterior differential $d$ in the form
$$d=d_0+d_1+t_0+t_1$$
with bi-degrees: $d_0 \sim (1,0)$, $d_1 \sim (0,1)$, $t_0 \sim (-1,2)$, $t_1 \sim (2,-1)$.

It follows from $d^2=0$ that
\begin{eqnarray}
t_0^2 &=& 0 \\
d_1t_0+t_0d_1 &=& 0 \\
d_1^2 +d_0t_0+t_0d_0 &=& 0
\end{eqnarray}
This means, in particular, that $t_0$ can be considered as a differential; corresponding cohomology groups will be denoted by $H^{p,q}_t.$ The differential $t_0$ can be identified with the differential
$$\gamma^m_{\alpha\beta}t^\alpha t^\beta \frac{\partial}{\partial C^m}$$ appearing in the calculation of cohomology  group of  the supersymmetry algebra $susy$; hence the cohomology groups $H^{p,q}_t$  coincide with graded components of cohomology groups of the Lie superalgebra $susy.$

Cohomology groups of the Lie superalgebra $susy$ were calculated in \cite {MSX}, this allows us to compute the groups $H^{p,q}_t.$ However, this is not the end of the story: the operator $d_1$ induces a differential on $H^{p,q}_t$; corresponding homology groups are denoted by $H^{p,q}_s.$  (See  \cite {BH}.) We would like to calculate these groups. We give complete answers for zero momentum (independent of coordinate variables $x^m$) part of   $H^{p,q}_s.$ (It will be denoted by  $\mathcal{H}^{p,q}_s.)$

Ordinary supersymmetry superspace, which we used in the above construction, can be replaces by a superspace that supports extended supersymmetry.   We analyze the case of $N=2$ supersymmetry in ten-dimensional space. Notice that in this case the group  $\mathcal{H}^{0,q}_s$ can be interpreted
as the cohomology of the differential $t_L\frac{\partial}{\partial \theta_L}+t_R\frac{\partial}{\partial \theta_R}$ where  the spinors $t_L,t_R$ obey the relaxed pure spinor condition
$t_L\gamma ^mt_L+t_R\gamma ^mt_R=0$.  N. Berkovits informed us that the problem of calculation of this cohomology arises in pure spinor formalism of ten-dimensional supergravity;   Section 3.2 answers his question. The same cohomology appears also in \cite {Movshev:11D}

We would like to thank N. Berkovits and A. Mikhailov for thought provoking discussions.
	\section {Groups $H^{p,q}_t$ and $H^{p,q}_s$ : general results \label{Sec:HtHsgeneral}}

	In \cite{MSX} we have calculated the cohomology groups  $H^p(susy)$ as graded modules in polynomial algebra $\mathbb{C}[t]$ generated by $t^{\alpha}$.  These groups can  can be considered also as $Aut$-representations where $Aut$ stands for the group of automorphisms of $susy$; we have found the action of $Aut$ on them.
	
	The group $H^{p,q}_t$ is a graded component of $H^p(susy).$ More precisely, $H^{p,q}_t$ is a graded component of $H^p(susy)$ multiplied by the space of functions of $\theta^{\alpha}$ and $x^m$. (The differential $t_0$ has the same form as the differential in the definition of Lie algebra cohomology, but it acts on the space of functions depending on variables
$\theta^{\alpha}$ and $x^m$ in addition to the ghost variables $t^{\alpha}$ and $c^m$.)

To calculate the group $H^{p,q}_s$ we consider the differential \[d_s=t^{\alpha}\frac{\partial}{\partial \theta^{\alpha}}+t^\alpha\gamma^m_{\alpha\beta}\theta^\beta\frac{\partial}{\partial x^m}\] acting on  $H^p(susy)\otimes \Lambda (\theta)\otimes \mathbb{C}[x]$, where $\Lambda (\theta)$ is Grassmann algebra in $\theta^{\alpha}$.

As a first step in this calculation we will simplify this differential omitting the second term. The cohomology corresponding to this differential $d^0_s$ will be denoted by ${\mathcal{H}^{p,q}_s}$.(In other words, we consider the differential on functions that do not depend on $x$. This corresponds to zero momentum.)

First of all we notice that the differential 	$d^0_s=t^{\alpha}\frac{\partial}{\partial \theta^{\alpha}}$ acting on
$\mathbb{C}[t] \otimes\Lambda [\theta]$ is acyclic (only constants give non-trivial cohomology classes). This follows immediately from the commutation relation $hd^0_s+d^0_sh=N$ where $h=\theta^{\alpha}\frac{\partial}{\partial t^{\alpha}}$ and $N$ counts the number of $\theta$'s and $t$'s (denoted by $\overline{\theta^\alpha}$ and $\overline{t^\alpha}$ respectively): $N=\displaystyle\sum_\alpha(\overline{\theta^\alpha}+\overline{t^\alpha}).$
It follows that  a cocycle with $N\neq 0$ is a coboundary (if $d^0_sx=0$ then $x=d^0_s(N(x)^{-1}hx)$). The same is true if we replace $\mathbb{C}[t]$ by a free module $\mu\otimes \mathbb{C}[t]$. The cohomology $H^p(susy)$ is not a free module, but it has a free resolution $\Gamma^p=\sum \Gamma^p _i=\sum \mu^ p_i\otimes \mathbb{C}[t]$ constructed in \cite {MSX}. In other words there exists a graded module $\Gamma$ with differential $\delta$ acting from $\Gamma_i$ into $\Gamma_{i-1}$; this differential is acyclic in all degrees except zero,  the only non-trivial homology is isomorphic to $H^p(susy).$

Let us  consider now a differential $D=d^0_s+\delta$  acting in $\Gamma^p\otimes \Lambda [\theta].$  (we omit the dependence of $p$ in these notations). We represent a cocycle $v$ of this differential in the form $v=\sum v_k$ where $v_k\in \Gamma _k\otimes \Lambda [\theta]$. The equation $Dv=0$ can be represented in the form $d^0_sv_k+\delta v_{k+1}=0.$  Denote the last non-zero term of the sequence $v_k$ by $v_m.$ Then $d^0_sv_m=0$, hence $v_m$ specifies a cohomology class of the differential $d^0_s.$ We see that possible $v_m$ are labelled by elements of $\mu^p_m.$ (Changing $v_m$ inside of its cohomology class we do not change the cohomology class of $v.$) We conclude that the cohomology of $D$ is isomorphic to the
direct sum of $\mu^p_m.$ Knowing $v_m$ we can find all $v_k$ inductively using the formula
\begin{equation}
\label{e}
v_k=-N(v_{k+1})^{-1}h\delta v_{k+1}.
\end{equation}

 From the other side we can solve the equations for $v_k$ starting with $v_0$. Then we can identify the cohomology of $D$ with the cohomology of  $H^p(susy)\otimes \Lambda [\theta]$ with respect to the differential $d^0_s,$ i.e. with  the direct sum of groups  $\mathcal{H}^{p,q}_s$ for all $q$. It follows that this direct sum is isomorphic to $A^p=\sum_m \mu^p_m$

To find the part of $A^p$ corresponding to the group  $\mathcal{H}^{p,q}_s$ we should use (\ref {e}) and the $t$-degrees of  elements of $\mu^p_m$ (this information can be found in \cite {MSX}).  It is easy to check the following statement: an element  of $ \mu^p_m$ having $t$-degree $r$ corresponds to an element of $\mathcal{H}^{p,q}_s$ where
\begin{equation}
\label{pq}
q=r -k.
\end{equation}
In other words the $t$-degree of this element is $r -k$.
This follows from the  remark that the operator $h$  decreases the $t$-degree by $1$ . It increases $\theta$-degree by $1$ , hence the $\theta$-degree of this element is $k.$

Let us denote by $\mu^p_{k,r}$ the part of $\mu^p_k$ consisting of elements of $t$-degree $r.$ Then it  follows from the above considerations that
\begin{equation}
\label{0}
 \mathcal{H}^{p,q}_s=\sum_k \mu^p_{k,q+k}
\end{equation}
More precisely, $\mu^p_{k,q+k}$ is isomorphic  $ (\mathcal{H}^{p,q}_s)_k$ (to the subgroup of  $ \mathcal{H}^{p,q}_s$ consisting of elements of $\theta$-degree $k).$

More rigorous treatment of the above calculations should be based on application of spectral sequences of bicomplex.  Namely, the cohomology of $D=d^0_s+\delta$ can be calculated by means
of spectral sequence starting with $H(H(d^0_s),\delta)$ or by means of spectral sequence starting with
$H(H(\delta),d^0_s).$ Both spectral sequences collapse, hence the cohomology of $D$ is equal to
 $$H(H(d^0_s),\delta)=H(H(\delta),d^0_s).$$
 By the definition of resolution $H(\delta)=H^p(susy)$, hence the RHS is a direct sum of   $\mathcal{H}^{p,q}_s$. The LHS is equal to $\sum \mu^p_m.$(The group $H(d^0_s)$ is isomorphic to $\sum \mu^p_m$
and the differential $\delta$ acts trivially on this group.)

Let us discuss the calculation of $H^{p,q}_s.$ 	

First of all we notice that  the differential $d_s$ acting on  $H^p(susy)\otimes \Lambda (\theta)\otimes \mathbb{C}[x]$ anti-commutes  with operators $Q_{\alpha}=\frac{\partial }{\partial \theta ^{\alpha}}-\gamma^m_{\alpha\beta}\theta^\beta\frac{\partial}{\partial x^m}$ generating supersymmetry Lie algebra. This means that  $H^{p,q}_s$ can be considered as a representation of this Lie algebra (moreover, it is a representation of super Poincare Lie algebra).

The differential $d_s$  can be represented in the form $d_s=d^0_s+d^1_s$, where $d^1_s=t^\alpha\gamma^m_{\alpha\beta}\theta^\beta\frac{\partial}{\partial x^m}$.
We can apply spectral sequence of bicomplex to calculate its cohomology. However, we will use more pedestrian approach considering the second summand in $d_s$ as perturbation,
In other words, we are writing $d_s$ as $d^0_s+\epsilon d^1_s$ and looking for a solution of the equation $d_sa=0$ in the form $a=a_0+\epsilon a_1+\epsilon^2 a_2+\cdots$ We obtain
$$d^0_sa_0=0,$$
$$d^0_sa_1+d^1_sa_0=0,$$
$$d^0_sa_2+d^1_sa_1=0,$$
$$\cdots$$

The operator $d^1_s$ descends to a differential on the cohomology of the operator $d^0_s$. In the first approximation the cohomology of $d_s$ is equal to the cohomology of $d^1_s$ in the cohomology of $d^0_s$.  This is clear from the  above formulas ( the cohomology class of $a_0$  in the cohomology of $d^0_s$ is a cocycle of $d^1_s).$  It follows  that in the first approximation  the homology $H^{p,q}_s$ is equal to $H(\mathcal{H}_s^{p,q}\otimes \mathbb{C}[x],d^1_s).$

The differential $$d^1_s:\mathcal{H}_s^{p,q}\otimes \mathbb{C}[x]\to \mathcal{H}_s^{p,q+1}\otimes \mathbb{C}[x]$$
has the form $d^1_sf(x)=R\nabla f(x)$ where $R$ denotes an intertwiner (an $Aut$-invariant map)
$R:\mathcal{H}^{p,q}_s \otimes V\to \mathcal{H}^{p,q+1}_s.$ Here $\nabla f$ stands for the gradient $\frac{\partial f}{\partial x^m}$, and $R$ is considered as a map of $Aut$-modules where $Aut$ stands for the automorphism group of $susy$, the letter $V$ denotes the vector representation of $Aut.$
Using (\ref {0}) we can interpret $R$ as intertwiner
$$R:\mu^p_{k,r}\otimes V\to \mu^p_{k+1,r+2}$$
In many cases this intertwiner is unique (up to a constant factor); we can use the LiE program to establish this fact.

In the language of spectral sequences the first approximation is the $E_2$ term.  There is a natural differential on $E_2$; its cohomology $E_3$ can be regarded as the next approximation. The calculation of the differential on $E_2$ can be based on the remark that the super Poincare group acts on homology.

\section {Groups $H^{p,q}_t$ and $\mathcal{H}^{p,q}_s$ in ten-dimensional space}

 \subsection {N=1 superspace in 10D\label{H:10D_N=1}}

Groups  $H^{p,q}_t$ were calculated in \cite {MSX} . The direct sum $\sum _q H^{p,q}_t$ can be considered as $\mathbb{C}[t]$ -module; the resolutions of these modules also were calculated in this  paper. We denote by $\mu^p_k$ the $k$-th term of the resolution and by $\mu^p_{k,r}$ the part of $\mu^p_k$ consisting of elements of $t$-degree $r.$ We  use a shorthand notation $[i_1,\dots, i_{[n/2]}]$ for  an irreducible representation $V$ of $SO(n)$ where  $i_1,\dots, i_{[n/2]}$ stand for coordinates of the  highest weight of $V$.

Using (\ref{0}) we obtain the groups $\mathcal{H}^{p,q}_s$:
\begin{itemize}
\item $p=0$,
$$\mathcal{H}^{0,0}_s=\displaystyle\sum_k \mu^0_{k,k}=\mu^0_{0,0}=[0,0,0,0,0],$$
$$\mathcal{H}^{0,1}_s=\displaystyle\sum_k \mu^0_{k,k+1}=\mu^0_{1,2}+\mu^0_{2,3}=[1,0,0,0,0]+[0,0,0,1,0],$$
$$\mathcal{H}^{0,2}_s=\displaystyle\sum_k \mu^0_{k,k+2}=\mu^0_{3,5}+\mu^0_{4,6}=[0,0,0,0,1]+[1,0,0,0,0],$$
$$\mathcal{H}^{0,3}_s=\displaystyle\sum_k \mu^0_{k,k+3}=\mu^0_{5,8}=[0,0,0,0,0];$$

\item $p=1$,
$$\mathcal{H}^{1,1}_s=\displaystyle\sum_k \mu^1_{k,k+1}=\mu^1_{0,1}+\mu^1_{1,2}=[0,0,0,1,0]+[0,1,0,0,0],$$
$$\mathcal{H}^{1,2}_s=\displaystyle\sum_{k=1}^{16} \mu^1_{k,k+2}=[0,0,0,0,1]+2[0,0,1,0,0] +[1,0,0,0,0]+[0,0,0,1,0] +[0,1,0,1,0]+$$
  $$\phantom{\mathcal{H}^{1,2}_s=} +[1,0,0,0,1]+[0,2,0,0,0] +[1,0,0,2,0] +[2,0,0,0,0]+[0,0,0,3,0] +[1,1,0,1,0]+$$
  $$\phantom{\mathcal{H}^{1,2}_s=} +[0,1,0,2,0]+[2,0,1,0,0]+[1,0,1,1,0]+[3,0,0,0,1]+[0,0,2,0,0]+[2,0,0,1,1]+$$
  $$\phantom{\mathcal{H}^{1,2}_s=} +[4,0,0,0,0]+[1,0,1,0,1]+[3,0,0,1,0]+[0,1,0,0,2]+[2,0,1,0,0]+[0,0,0,0,3]+$$
  $$\phantom{\mathcal{H}^{1,2}_s=} +[1,1,0,0,1]+[0,2,0,0,0]+[1,0,0,0,2]+[0,1,0,0,1]+[0,0,1,0,0]+[0,0,0,1,0]+$$
  $$\phantom{\mathcal{H}^{1,2}_s=} +[0,0,0,0,0],$$
$$\mathcal{H}^{1,3}_s=\displaystyle\sum_k \mu^1_{k,k+3}=\mu^1_{4,7}+\mu^1_{5,8}=[0,0,0,0,1]+[1,0,0,0,0];$$

\item $p=2$,
$$\mathcal{H}^{2,2}_s=\displaystyle\sum_{k=0}^{14} \mu^2_{k,k+2}=[0,0,1,0,0]+[0,0,0,1,0] +[0,1,0,1,0] +[1,0,0,0,1]+[0,0,0,0,0]+$$
  $$\phantom{\mathcal{H}^{2,2}_s=} +[0,0,0,1,1] +[0,1,0,0,0]+[0,2,0,0,0] +[1,0,0,2,0]+ [2,0,0,0,0]+[0,0,0,3,0]+$$
  $$\phantom{\mathcal{H}^{2,2}_s=} +[1,0,0,1,0] +[1,1,0,1,0]+[0,1,0,2,0] +[2,0,1,0,0]+[1,0,1,1,0] +[3,0,0,0,1]+$$
  $$\phantom{\mathcal{H}^{2,2}_s=} +[0,0,2,0,0] +[2,0,0,1,1]+[4,0,0,0,0] +[1,0,1,0,1]+[3,0,0,1,0]+[0,1,0,0,2]+$$
  $$\phantom{\mathcal{H}^{2,2}_s=} +[2,0,1,0,0] +[0,0,0,0,3]+[1,1,0,0,1] +[0,2,0,0,0]+[1,0,0,0,2]+[0,1,0,0,1]+$$
  $$\phantom{\mathcal{H}^{2,2}_s=} +[0,0,1,0,0]+[0,0,0,1,0]+[0,0,0,0,0],$$
$$\mathcal{H}^{2,3}_s=\displaystyle\sum_k \mu^2_{k,k+3}=\mu^2_{3,6}+\mu^2_{4,7}+\mu^2_{5,8}=[0,0,0,0,2] +[1,0,0,0,0]+[0,0,0,1,0]+$$
  $$\phantom{\mathcal{H}^{2,3}_s=} +[1,0,0,0,1]+[0,1,0,0,0];$$

\item $p=3$,
$$\mathcal{H}^{3,2}_s=\displaystyle\sum_k \mu^3_{k,k+2}=\mu^3_{0,2}+\mu^3_{1,3}+\mu^3_{2,4}=[0,1,0,0,0]+[0,0,0,0,1] +[1,0,0,1,0]+$$
  $$\phantom{\mathcal{H}^{3,2}_s=} +[0,0,0,2,0] +[1,0,0,0,0],$$
$$\mathcal{H}^{3,3}_s=\displaystyle\sum_k \mu^3_{k,k+3}=\mu^3_{2,5}+\mu^3_{3,6}+\mu^3_{4,7}=[1,0,0,0,1]+[0,0,0,0,0] +[0,0,0,1,1]+$$
  $$\phantom{\mathcal{H}^{3,3}_s=} +[0,1,0,0,0]+[2,0,0,0,0]+[0,0,0,0,1]+[1,0,0,1,0],$$
$$\mathcal{H}^{3,4}_s=\displaystyle\sum_k \mu^3_{k,k+4}=\mu^3_{5,9}+\mu^3_{6,10}=[0,0,0,1,0]+[0,0,0,0,0];$$

\item $p=4$,
$$\mathcal{H}^{4,2}_s=\displaystyle\sum_k \mu^4_{k,k+2}=\mu^4_{0,2}+\mu^4_{1,3}=[1,0,0,0,0]+[0,0,0,1,0],$$
$$\mathcal{H}^{4,3}_s=\displaystyle\sum_k \mu^4_{k,k+3}=\mu^4_{1,4}+\mu^4_{2,5}+\mu^4_{3,6}=[2,0,0,0,0]+[0,0,0,0,1] +[1,0,0,1,0]+$$
  $$\phantom{\mathcal{H}^{4,3}_s=} +[0,0,1,0,0] +[1,0,0,0,0],$$
$$\mathcal{H}^{4,4}_s=\displaystyle\sum_k \mu^4_{k,k+4}=\mu^4_{4,8}+\mu^4_{5,9}=[0,0,0,0,0] +[0,1,0,0,0]+[0,0,0,0,1];$$

\item $p=5$,
$$\mathcal{H}^{5,2}_s=\displaystyle\sum_k \mu^5_{k,k+2}=\mu^5_{0,2}=[0,0,0,0,0],$$
$$\mathcal{H}^{5,3}_s=\displaystyle\sum_k \mu^5_{k,k+3}=\mu^5_{1,4}+\mu^5_{2,5}=[1,0,0,0,0]+[0,0,0,1,0],$$
$$\mathcal{H}^{5,4}_s=\displaystyle\sum_k \mu^5_{k,k+4}=\mu^5_{3,7}+\mu^5_{4,8}=[0,0,0,0,1]+[1,0,0,0,0],$$
$$\mathcal{H}^{5,5}_s=\displaystyle\sum_k \mu^5_{k,k+5}=\mu^5_{5,10}=[0,0,0,0,0].$$

\end{itemize}
all other groups vanish. The results above are verified by $SO$ character-valued  Euler characteristics 
\beq \label{eulerchar:Hs}
\chi_{p,q+k}:=\sum_k [(\mathcal{H}^{p,q}_s)_k] (-1)^k=\sum_k [H^{p,q}_t\otimes \Lambda^k(\theta)] (-1)^k
\eeq
where $k$ corresponds to the degree of $\theta$. We use a shorthand notation $[V]$ for the character $\chi_V(g)$ of the representation $V$. The formula (\ref{eulerchar:Hs}) reflects the idea  that the character-valued Euler characteristics should be the same for the chain complex $C^{p,q,k}:=H^{p,q}_t\otimes \Lambda^k(\theta)$ and  its  cohomology groups $(\mathcal{H}^{p,q}_s)_k$.
Since their associated differential is $d^0_s=t^{\alpha}\frac{\partial}{\partial \theta^{\alpha}}$, we can do alternated summation by varying the degree of $\theta$, while the degrees of $p$ and $q+k$ are invariant.

Let us consider in more detail the case of zero-dimensional cohomology in ten-dimensional space. These calculations can be compared with computations in \cite {B}. Higher dimensional cohomology can be analyzed in similar way.

It is easy to see that  $H^0(susy)=\sum _qH^{0,q}_t$ is  the space of polynomial functions on pure spinors (in other words, this is a quotient of $\mathbb{C}[t]$ with respect to the ideal generated by  $\gamma^m_{\alpha \beta}t^{\alpha}t^{ \beta}).$

According to \cite {MSX}  the  resolution of  $H^0(susy)$ has the form $\Gamma_k=\mu_k\otimes \mathbb{C}[t]$ where

$$\mu_0=[0,0,0,0,0], \dim(\mu_0)=1, \deg(\mu_0)=0; $$
$$\mu_1=[1,0,0,0,0], \dim(\mu_1)=10, \deg(\mu_1)=2; $$
$$\mu_2=[0,0,0,1,0], \dim(\mu_2)=16, \deg(\mu_2)=3; $$
$$\mu_3=[0,0,0,0,1], \dim(\mu_3)=16, \deg(\mu_3)=5; $$
$$\mu_4=[1,0,0,0,0], \dim(\mu_4)=10, \deg(\mu_4)=6; $$
$$\mu_5=[0,0,0,0,0], \dim(\mu_5)=1, \deg(\mu_5)=8 $$

where $\mu_k$ are considered  as representations of the group $Aut= SO(10).$

Here the degree is calculated with respect to $t$, the differential $\delta=\sum\delta_k$ preserves the degree (this means, for example, that $\delta_5$ is a multiplication by a quadratic (with respect to $t$) polynomial).

We will denote the elements of $\mu^0_{0,0},\cdots,\mu^0_{5,8}$ by $a,c^m,g_{\alpha},k^{\alpha},s_m,u$ respectively.
These elements can be identified with elements of $\mathcal{H}^{p,q}_s$  by means of (\ref {e}).
Corresponding  elements of  $\mathcal{H}^{p,q}_s$  have degrees $$(0,0), (1,1), (1,2), (2,3), (2,4), (3,5),$$ where the first number stands for the $t$-degree and the second number denotes the $\theta$-degree (this follows from Eq.~(\ref{pq})).

There are two ways to calculate these elements explicitly (up to a constant factor). First of all the $SO(10)$-invariance of the resolution permits us to guess the expression for $\delta_k$. It is easy to see that the maps
$$\delta_1 c^m=\gamma^m_{\alpha \beta}t^{\alpha }t^{\beta},$$
$$\delta_2 g_{\alpha}=\gamma^m_{\alpha \beta}t^{\beta},$$
$$\delta_3 k_{\alpha}=\gamma^m_{\alpha_1 \beta_1}\gamma^m_{\alpha_2 \beta_2}t^{\alpha_1}t^{\beta_1},$$
$$\delta_4 s_{m}=\gamma^m_{\alpha \beta}t^{\beta},$$
$$\delta_5 u=\gamma^m_{\alpha \beta}t^{\alpha }t^{\beta}.$$
specify an $SO(10)$-invariant complex. From another side using LiE code one can  check that  $SO(10)$-invariance specifies this choice of differentials uniquely (up to a constant factor). Knowing $\delta_k$ we can calculate the elements  belonging to the cohomology classes in  $\mathcal{H}^{p,q}_s$  using (\ref {e}). We obtain
the elements
$$[\gamma^m_{\alpha\beta}t^{\alpha}\theta^{\beta}] \in \mathcal{H}^{0,1}_s,$$
$$[(t\gamma_m \theta)(\gamma^m\theta)_\alpha]  \in \mathcal{H}^{0,1}_s,$$
$$[(t\gamma_m \theta)(t\gamma^n \theta)(\theta\gamma_{mn})^\alpha] \in \mathcal{H}^{0,2}_s,$$
$$[(t\gamma_m \theta)(t\gamma^n \theta)(\theta\gamma_{mnp}\theta)] \in \mathcal{H}^{0,2}_s,$$
$$[(t\gamma_m \theta)(t\gamma^n \theta)(t\gamma^p \theta)(\theta\gamma_{mnp}\theta)]  \in \mathcal{H}^{0,3}_s.$$
 Another way is based on the remark that we know the degrees and transformation properties of these elements. Again using LiE code we can find all possible answers.

\subsection {$N=1$ superspace in $11D$}
In out previous paper~\cite{MSX}, we already obtained the cohomology groups of $H_t^{p,q}$ in $11D$. Then based on the Euler characteristics of $\mu_i$, namely the Eq.(109) in our paper~\cite{MSX},
we could find $\mu_k$ by matching the grading in the LHS and RHS of the equation (assuming there is no highest weight vector representation with the same degree appearing in two $\mu_k$'s). The results for the case when $p=0$ are verified with Movshev's result~\cite{Movshev:11D}.

\begin{itemize}
\item $p=0,$
$$\mu^0_0=\mu^0_{0,0}=[0,0,0,0,0], \dim(\mu^0_{0,0})=1,$$ 
$$\mu^0_1=\mu^0_{1,2}=[1,0,0,0,0], \dim(\mu^0_{1,2})=11,$$ 
$$\mu^0_2=\mu^0_{2,4}=[0,1,0,0,0]+[1,0,0,0,0], \dim(\mu^0_{2,4})=66,$$ 
$$\mu^0_3=\mu^0_{3,5}+\mu^0_{3,6}, $$
$$\mu^0_{3,5}=[0,0,0,0,1], \dim(\mu^0_{3,5})=32,$$ 
$$\mu^0_{3,6}=[0,0,0,0,0]+[0,0,1,0,0]+[2,0,0,0,0], \dim(\mu^0_{3,6})=231,$$ 
$$\mu^0_4=\mu^0_{4,7}=[0,0,0,0,1]+[1,0,0,0,1], \dim(\mu^0_{4,7})=352,$$ 
$$\mu^0_5=\mu^0_{5,9}=[0,0,0,0,1]+[1,0,0,0,1], \dim(\mu^0_{5,9})=352,$$ 
$$\mu^0_6=\mu^0_{6,10}+\mu^0_{6,11}, $$
$$\mu^0_{6,10}=[0,0,0,0,0]+[0,0,1,0,0]+[2,0,0,0,0], \dim(\mu^0_{6,10})=231,$$ 
$$\mu^0_{6,11}=[0,0,0,0,1], \dim(\mu^0_{6,11})=32,$$ 
$$\mu^0_7=\mu^0_{7,12}=[0,1,0,0,0]+[1,0,0,0,0], \dim(\mu^0_{7,12})=66,$$ 
$$\mu^0_8=\mu^0_{8,14}=[1,0,0,0,0], \dim(\mu^0_{8,14})=11,$$ 
$$\mu^0_9=\mu^0_{9,16}=[0,0,0,0,0], \dim(\mu^0_{9,16})=1.$$ 
%
%
%
%

\item $p=1,$
$$\mu^1_0=\mu^1_{0,4}=[1,0,0,0,0], \dim(\mu^1_{0,4})=11,$$ 
$$\mu^1_1=\mu^1_{1,5}+\mu^1_{1,6}, $$
$$\mu^1_{1,5}=[0,0,0,0,1], \dim(\mu^1_{1,5})=32,$$ 
$$\mu^1_{1,6}=[2,0,0,0,0], \dim(\mu^1_{1,6})=65,$$ 
$$\mu^1_2=\mu^1_{2,7}=[0,0,0,0,1]+[1,0,0,0,1], \dim(\mu^1_{2,7})=352,$$ 
$$\mu^1_3=\mu^1_{3,8}+\mu^1_{3,9}, $$
$$\mu^1_{3,8}=[0,0,0,1,0]+[1,0,0,0,0], \dim(\mu^1_{3,8})=341,$$ 
$$\mu^1_{3,9}=[0,0,0,0,1]+[1,0,0,0,1], \dim(\mu^1_{3,9})=352,$$ 
$$\mu^1_4=\mu^1_{4,10}=[0,0,0,0,0]+[0,0,0,0,2]+[0,0,1,0,0]+[0,1,0,0,0]+[1,0,0,0,0]+[2,0,0,0,0],$$ $$ \dim(\mu^1_{4,10})=759,$$ 
$$\mu^1_5=\mu^1_{5,12}=[0,0,0,0,0]+[0,0,0,0,2]+[0,0,1,0,0]+[0,1,0,0,0]+[1,0,0,0,0]+[2,0,0,0,0],$$ $$\dim(\mu^1_{5,12})=759,$$ 
$$\mu^1_6=\mu^1_{6,13}+\mu^1_{6,14}, $$
$$\mu^1_{6,13}=[0,0,0,0,1]+[1,0,0,0,1], \dim(\mu^1_{6,13})=352,$$ 
$$\mu^1_{6,14}=[0,0,0,1,0]+[1,0,0,0,0], \dim(\mu^1_{6,14})=341,$$ 
$$\mu^1_7=\mu^1_{7,15}=[0,0,0,0,1]+[1,0,0,0,1], \dim(\mu^1_{7,15})=352,$$ 
$$\mu^1_8=\mu^1_{8,16}+\mu^1_{8,17}, $$
$$\mu^1_{8,16}=[2,0,0,0,0], \dim(\mu^1_{8,16})=65,$$ 
$$\mu^1_{8,17}=[0,0,0,0,1], \dim(\mu^1_{8,17})=32,$$ 
$$\mu^1_9=\mu^1_{9,18}=[1,0,0,0,0], \dim(\mu^1_{9,18})=11.$$ 
%
%
%
%
%
%
%
%
%
%
\item $p=2,$
$$\mu^2_0=\mu^2_{0,2}=[0,0,0,0,0], \dim(\mu^2_{0,2})=1,$$ 
$$\mu^2_1=\mu^2_{1,4}=[1,0,0,0,0], \dim(\mu^2_{1,4})=11,$$ 
$$\mu^2_2=\mu^2_{2,6}=[0,1,0,0,0]+[1,0,0,0,0], \dim(\mu^2_{2,6})=66,$$ 
$$\mu^2_3=\mu^2_{3,7}+\mu^2_{3,8}, $$
$$\mu^2_{3,7}=[0,0,0,0,1], \dim(\mu^2_{3,7})=32,$$ 
$$\mu^2_{3,8}=[0,0,0,0,0]+[0,0,1,0,0]+[2,0,0,0,0], \dim(\mu^2_{3,8})=231,$$ 
$$\mu^2_4=\mu^2_{4,9}=[0,0,0,0,1]+[1,0,0,0,1], \dim(\mu^2_{4,9})=352,$$ 
$$\mu^2_5=\mu^2_{5,11}=[0,0,0,0,1]+[1,0,0,0,1], \dim(\mu^2_{5,11})=352,$$ 
$$\mu^2_6=\mu^2_{6,12}+\mu^2_{6,13}, $$
$$\mu^2_{6,12}=[0,0,0,0,0]+[0,0,1,0,0]+[2,0,0,0,0], \dim(\mu^2_{6,12})=231,$$ 
$$\mu^2_{6,13}=[0,0,0,0,1], \dim(\mu^2_{6,13})=32,$$ 
$$\mu^2_7=\mu^2_{7,14}=[0,1,0,0,0]+[1,0,0,0,0], \dim(\mu^2_{7,14})=66,$$ 
$$\mu^2_8=\mu^2_{8,16}=[1,0,0,0,0], \dim(\mu^2_{8,16})=11,$$ 
$$\mu^2_9=\mu^2_{9,18}=[0,0,0,0,0], \dim(\mu^2_{9,18})=1.$$ 
%
%
%
%
%
%
%
%
%
\end{itemize}

Now following the same procedure and notations of gradings as in Sec.~\ref{H:10D_N=1} and based on Eq.~(\ref{0}), we obtain the groups $\mathcal{H}^{p,q}_s:$
 \begin{itemize}
\item $p=0$,
$$\mathcal{H}^{0,0}_s=\mu^0_{0,0}=[0,0,0,0,0],$$
$$\mathcal{H}^{0,1}_s=\mu^0_{1,2}=[1,0,0,0,0],$$
$$\mathcal{H}^{0,2}_s=\mu^0_{2,4}+\mu^0_{3,5}=[0,1,0,0,0]+[1,0,0,0,0]+([0,0,0,0,1]),$$
$$\mathcal{H}^{0,3}_s=\mu^0_{3,6}+\mu^0_{4,7}=[0,0,0,0,0]+[0,0,1,0,0]+[2,0,0,0,0]+([0,0,0,0,1]+[1,0,0,0,1]),$$
$$\mathcal{H}^{0,4}_s=\mu^0_{5,9}+\mu^0_{6,10}=[0,0,0,0,1]+[1,0,0,0,1]+([0,0,0,0,0]+[0,0,1,0,0]+[2,0,0,0,0]),$$
$$\mathcal{H}^{0,5}_s=\mu^0_{6,11}+\mu^0_{7,12}=[0,0,0,0,1]+([0,1,0,0,0]+[1,0,0,0,0]),$$
$$\mathcal{H}^{0,6}_s=\mu^0_{8,14}=[1,0,0,0,0],$$
$$\mathcal{H}^{0,7}_s=\mu^0_{9,16}=[0,0,0,0,0];$$

\item $p=1$,
$$\mathcal{H}^{1,2}_s=\mu^1_{0,2}+\mu^1_{1,3}=[1,0,0,0,0]+([0,0,0,0,1]),$$
$$\mathcal{H}^{1,3}_s=\mu^1_{1,4}+\mu^1_{2,5}+\mu^1_{3,6}=[2,0,0,0,0]+([0,0,0,0,1]+[1,0,0,0,1])+([0,0,0,1,0]+[1,0,0,0,0]),$$
$$\mathcal{H}^{1,4}_s=\mu^1_{3,7}+\mu^1_{4,8}=[0,0,0,0,1]+[1,0,0,0,1]+([0,0,0,0,0]+[0,0,0,0,2]+[0,0,1,0,0]+$$
$$\phantom{\mathcal{H}^{1,4}_s=}+[0,1,0,0,0]+[1,0,0,0,0]+[2,0,0,0,0]),$$
$$\mathcal{H}^{1,5}_s=\mu^1_{5,10}+\mu^1_{6,11}=[0,0,0,0,0]+[0,0,0,0,2]+[0,0,1,0,0]+[0,1,0,0,0]+[1,0,0,0,0]+$$
$$\phantom{\mathcal{H}^{1,5}_s=}+[2,0,0,0,0]+([0,0,0,0,1]+[1,0,0,0,1]),$$
$$\mathcal{H}^{1,6}_s=\mu^1_{6,12}+\mu^1_{7,13}+\mu^1_{8,14}=[0,0,0,1,0]+[1,0,0,0,0]+([0,0,0,0,1]+[1,0,0,0,1])+([2,0,0,0,0]),$$
$$\mathcal{H}^{1,7}_s=\mu^1_{8,15}+\mu^1_{9,16}=[0,0,0,0,1]+([1,0,0,0,0]);$$

\item $p=2$,
$$\mathcal{H}^{2,2}_s=\mu^2_{0,2}=[0,0,0,0,0],$$
$$\mathcal{H}^{2,3}_s=\mu^2_{1,4}=[1,0,0,0,0],$$
$$\mathcal{H}^{2,4}_s=\mu^2_{2,6}+\mu^2_{3,7}=[0,1,0,0,0]+[1,0,0,0,0]+([0,0,0,0,1]),$$
$$\mathcal{H}^{2,5}_s=\mu^2_{3,8}+\mu^2_{4,9}=[0,0,0,0,0]+[0,0,1,0,0]+[2,0,0,0,0]+([0,0,0,0,1]+[1,0,0,0,1]),$$
$$\mathcal{H}^{2,6}_s=\mu^2_{5,11}+\mu^2_{6,12}=[0,0,0,0,1]+[1,0,0,0,1]+([0,0,0,0,0]+[0,0,1,0,0]+[2,0,0,0,0]),$$
$$\mathcal{H}^{2,7}_s=\mu^2_{6,13}+\mu^2_{7,14}=[0,0,0,0,1]+([0,1,0,0,0]+[1,0,0,0,0]),$$
$$\mathcal{H}^{2,8}_s=\mu^2_{8,16}=[1,0,0,0,0],$$
$$\mathcal{H}^{2,9}_s=\mu^2_{9,18}=[0,0,0,0,0].$$

\end{itemize}
all other cohomology groups vanish. The results above are verified by Euler characteristics defined by Eq.~\ref{eulerchar:Hs}.

 \subsection {$N=2$ superspace in $10D$}

 In this superspace we have coordinates $z^M \sim (x^m,  \theta _L,\theta_R)$ (vector and two spinors).
 We take a basis in the space of 1-forms given by formulas $E^m=dx^m+\Gamma^m_{\alpha\beta}\theta_L^\alpha d\theta_L^\beta+\Gamma^m_{\alpha\beta}\theta_R^\alpha d\theta_R^\beta, E_L^{\alpha}=d\theta_L^\alpha, E_R^{\alpha}=d\theta_R^\alpha. $
 The differential that is used to calculate the cohomology $H_t$ (= the cohomology of $susy_2$, i.e. of the Lie algebra of N=2 supersymmetry) has the form
 $$d=\gamma^m_{\alpha \beta}t_L^{\alpha} t_L^{\beta} \frac{\partial}{\partial C^m} +\gamma^m_{\alpha\beta}t_R^{\alpha} t_R^{\beta} \frac{\partial}{\partial C^m}$$
It acts on the space of functions of commuting variables $t_L^{\alpha},t_R^{\alpha}$ and anticommuting variables $C^m$. It is useful to  make a change of variables $t_+=t_L+it_R,t_-=t_L-it_R,$, then the differential takes the form
$$d=2\gamma^m_{\alpha\beta}t_+^\alpha t_-^\beta\frac{\partial}{\partial C^m}$$
The automorphism group of $N=2,D=10$  $susy$ is $SO(10)\times GL(1)$. The variables $t_+,t_-$ are $SO(10)$ spinors having the weight $+1$ and $-1$ with respect to $GL(1)$, the variables $C_m$  constitute a vector of weight $0$.

The cohomology $H_t$ has three gradings: with respect to the number of  ghosts (of $C$-variables, and denoted by $\overline{C}$), to the number of $t_+$-variables (denoted by $\overline{t_+}$) and to the number of $t_-$-variables (denoted by $\overline{t_-}$). Hence we could denote its components by $H_t^{\overline{C},\overline{t_+},\overline{t_-}}$. Sometimes instead of $t_+$ and $t_-$ gradings, it is convenient to use $t$-grading $\overline t= \overline{t_+}+\overline{t_-}$ and $GL(1)$-grading $\overline{gl}=\overline{t_+}-\overline{t_-}.$

One can calculate it using the methods of \cite {MSX}.  Let us formulate the answers for  the case when the number of ghosts is 0 (zero-dimensional cohomology).

Again using {\tt Macaulay2} one can find the dimensions and gradings
of the components of the  resolution of the  zero-th cohomology considered as a module over the polynomial ring $\mathbb{C}[t_+,t_-]$. Using the LiE code we find the action of $SO(10)$ on these components.
We represent the resolution as an exact sequence $ ....\to\mu^0_k\to \mu^0_{k-1}\to ...$; the  components of $\mu^0_k$ are denoted by $\mu^0_{k, k+\overline{t}, \overline{gl} },$ where
$$\overline{t}=\overline{t_+}+\overline{t_-},\quad \overline{gl}=\overline{t_+}-\overline{t_-}.$$
\begin{itemize}
\item $p=0,$
$$\mu^0_0=\mu^0_{0,0,0}=[0,0,0,0,0], \dim(\mu^0_{0,0,0})=1, $$
$$\mu^0_1=\mu^0_{1,2,0}=[1,0,0,0,0], \dim(\mu^0_{1,2,0})=10, $$
$$\mu^0_2=\mu^0_{2,4,-2}+\mu^0_{2,4,2}+\mu^0_{2,4,0};$$
$$\mu^0_{2,4,-2}=[0,0,0,0,0], \dim(\mu^0_{2,4,-2})=1, $$
$$\mu^0_{2,4,2}=[0,0,0,0,0], \dim(\mu^0_{2,4,2})=1, $$
$$\mu^0_{2,4,0}=[0,1,0,0,0], \dim(\mu^0_{2,4,0})=45, $$
$$\mu^0_3=\mu^0_{3,6,-2}+\mu^0_{3,6,2}+\mu^0_{3,6,0};$$
$$\mu^0_{3,6,-2}=[1,0,0,0,0], \dim(\mu^0_{3,6,-2})=10, $$
$$\mu^0_{3,6,2}=[1,0,0,0,0], \dim(\mu^0_{3,6,2})=10, $$
$$\mu^0_{3,6,0}=[0,0,1,0,0] + [1,0,0,0,0], \dim(\mu^0_{3,6,0})=130, $$
$$\mu^0_4=\mu^0_{4,7,-1}+\mu^0_{4,7,1}+\mu^0_{4,8,-2}+\mu^0_{4,8,2}+\mu^0_{4,8,-4}+\mu^0_{4,8,4}+\mu^0_{4,8,0};$$
$$\mu^0_{4,7,-1}=[0,0,0,1,0], \dim(\mu^0_{4,7,-1})=16, $$
$$\mu^0_{4,7,1}=[0,0,0,1,0], \dim(\mu^0_{4,7,1})=16, $$
$$\mu^0_{4,8,-2}=[0,1,0,0,0], \dim(\mu^0_{4,8,-2})=45, $$
$$\mu^0_{4,8,2}=[0,1,0,0,0], \dim(\mu^0_{4,8,2})=45, $$
$$\mu^0_{4,8,-4}=[0,0,0,0,0], \dim(\mu^0_{4,8,-4})=1, $$
$$\mu^0_{4,8,4}=[0,0,0,0,0], \dim(\mu^0_{4,8,4})=1, $$
$$\mu^0_{4,8,0}=[0,0,0,0,0] +[0,0,0,1,1] +[2,0,0,0,0], \dim(\mu^0_{4,8,0})=265, $$
$$\mu^0_5=\mu^0_{5,9,-1}+\mu^0_{5,9,1}+\mu^0_{5,9,-3}+\mu^0_{5,9,3}+\mu^0_{5,10,0};$$
$$\mu^0_{5,9,-1}=[0,0,0,0,1] +[1,0,0,1,0], \dim(\mu^0_{5,9,-1})=160, $$
$$\mu^0_{5,9,1}=[0,0,0,0,1] +[1,0,0,1,0], \dim(\mu^0_{5,9,1})=160, $$
$$\mu^0_{5,9,-3}=[0,0,0,0,1], \dim(\mu^0_{5,9,-3})=16, $$
$$\mu^0_{5,9,3}=[0,0,0,0,1], \dim(\mu^0_{5,9,3})=16, $$
$$\mu^0_{5,10,0}=[0,0,0,0,2], \dim(\mu^0_{5,10,0})=126, $$
$$\mu^0_6=\mu^0_{6,11,-1}+\mu^0_{6,11,1}+\mu^0_{6,11,-3}+\mu^0_{6,11,3};$$
$$\mu^0_{6,11,-1}=[0,0,0,1,0] + [1,0,0,0,1], \dim(\mu^0_{6,11,-1})=160, $$
$$\mu^0_{6,11,1}=[0,0,0,1,0] + [1,0,0,0,1], \dim(\mu^0_{6,11,1})=160, $$
$$\mu^0_{6,11,-3}=[0,0,0,1,0], \dim(\mu^0_{6,11,-3})=16, $$
$$\mu^0_{6,11,3}=[0,0,0,1,0], \dim(\mu^0_{6,11,3})=16, $$
$$\mu^0_7=\mu^0_{7,13,-1}+\mu^0_{7,13,1}+\mu^0_{7,12,-2}+\mu^0_{7,12,2}+\mu^0_{7,12,-4}+\mu^0_{7,12,4}+\mu^0_{7,12,0};$$
$$\mu^0_{7,13,-1}=[0,0,0,0,1], \dim(\mu^0_{7,13,-1})=16, $$
$$\mu^0_{7,13,1}=[0,0,0,0,1], \dim(\mu^0_{7,13,1})=16, $$
$$\mu^0_{7,12,-2}=[0,1,0,0,0], \dim(\mu^0_{7,12,-2})=45, $$
$$\mu^0_{7,12,2}=[0,1,0,0,0], \dim(\mu^0_{7,12,2})=45, $$
$$\mu^0_{7,12,-4}=[0,0,0,0,0], \dim(\mu^0_{7,12,-4})=1, $$
$$\mu^0_{7,12,4}=[0,0,0,0,0], \dim(\mu^0_{7,12,4})=1, $$
$$\mu^0_{7,12,0}=[0,0,0,0,0] + [2,0,0,0,0], \dim(\mu^0_{7,12,0})=55, $$
$$\mu^0_8=\mu^0_{8,14,-2}+\mu^0_{8,14,2}+\mu^0_{8,14,0};$$
$$\mu^0_{8,14,-2}=[1,0,0,0,0], \dim(\mu^0_{8,14,-2})=10, $$
$$\mu^0_{8,14,2}=[1,0,0,0,0], \dim(\mu^0_{8,14,2})=10, $$
$$\mu^0_{8,14,0}=[1,0,0,0,0], \dim(\mu^0_{8,14,0})=10, $$
$$\mu^0_9=\mu^0_{9,16,-2}+\mu^0_{9,16,2};$$
$$\mu^0_{9,16,-2}=[0,0,0,0,0], \dim(\mu^0_{9,16,-2})=1, $$
$$\mu^0_{9,16,2}=[0,0,0,0,0], \dim(\mu^0_{9,16,2})=1, $$
\end{itemize}

Now we can use the methods of Section \ref{Sec:HtHsgeneral} to calculate  the group $\mathcal{H}_s^{0,\overline{t}, \overline{gl}}$.
 This group has an additional grading (the number of $\theta$'s); using the same considerations as in the proof of \ref {0}  we obtain the component with respect to this grading is given by the formula
\begin{equation}
\label{2}
(\mathcal{H}_s ^{0, \overline{t}, \overline{gl} })_k=\mu^0 _{k, \overline{t}+k, \overline{gl} }
\end{equation}

Comparing to the case of $N=1$ supersymmetry mentioned in Section \ref{Sec:HtHsgeneral}, we see that
$$d_s^0=t_L^\alpha \frac{\partial}{\partial \theta_L^\alpha} + t_R^\alpha \frac{\partial}{\partial \theta_R^\alpha}=t_+^\alpha \frac{\partial}{\partial \theta_+^\alpha} + t_-^\alpha \frac{\partial}{\partial \theta_-^\alpha},$$
$$h=\theta_+^\alpha \frac{\partial}{\partial t_+^\alpha} + \theta_-^\alpha \frac{\partial}{\partial t_-^\alpha},$$
\beq
\label{N:deg_t_theta}
N=\displaystyle\sum_\alpha(\overline{\theta_+^\alpha}+\overline{\theta_-^\alpha}+\overline{t_+^\alpha}+\overline{t_-^\alpha})
\eeq

To prove \ref {2} we use the fact that $h$ decreases the $t$-grading by $1$ and does not change the $GL(1)$ grading $\overline{gl}=\overline{t_+}-\overline{t_-}+\overline{\theta_+}-\overline{\theta_-}$ (we are assuming that $\theta_+,\theta_-$ have the $GL(1)$-gradings $+1$ and $-1$).


We obtain
$$(\mathcal{H}_s^{0,0,0})_0=\mu^0_{0,0,0}=[0,0,0,0,0],$$
$$(\mathcal{H}_s^{0,1,0})_1=\mu^0_{1,2,0}=[1,0,0,0,0],$$
$$(\mathcal{H}_s^{0,2,-2})_2=\mu^0_{2,4,-2}=[0,0,0,0,0],$$
$$(\mathcal{H}_s^{0,2,2})_2=\mu^0_{2,4,2}= [0,0,0,0,0],$$
$$(\mathcal{H}_s^{0,2,0})_2=\mu^0_{2,4,0}= [0,1,0,0,0],$$
%
$$(\mathcal{H}_s^{0,3,-2})_3=\mu^0_{3,6,-2}= [1,0,0,0,0],$$
$$(\mathcal{H}_s^{0,3,2})_3=\mu^0_{3,6,2}= [1,0,0,0,0],$$
$$(\mathcal{H}_s^{0,3,0})_3=\mu^0_{3,6,0}=[0,0,1,0,0] + [1,0,0,0,0],$$
%
$$(\mathcal{H}_s^{0,3,-1})_4=\mu^0_{4,7,-1}= [0,0,0,1,0],$$
$$(\mathcal{H}_s^{0,3,1})_4=\mu^0_{4,7,1}=[0,0,0,1,0],$$
$$(\mathcal{H}_s^{0,4,-2})_4=\mu^0_{4,8,-2}=[0,1,0,0,0],$$
$$(\mathcal{H}_s^{0,4,2})_4=\mu^0_{4,8,2}= [0,1,0,0,0],$$
$$(\mathcal{H}_s^{0,4,-4})_4=\mu^0_{4,8,-4}= [0,0,0,0,0],$$
$$(\mathcal{H}_s^{0,4,4})_4=\mu^0_{4,8,4}=[0,0,0,0,0],$$
$$(\mathcal{H}_s^{0,4,0})_4=\mu^0_{4,8,0}=[0,0,0,0,0] + [0,0,0,1,1] + [2,0,0,0,0],$$
$$(\mathcal{H}_s^{0,4,-1})_5=\mu^0_{5,9,-1}=[0,0,0,0,1] + [1,0,0,1,0],$$
$$(\mathcal{H}_s^{0,4,1})_5=\mu^0_{5,9,1}=[0,0,0,0,1] + [1,0,0,1,0],$$
$$(\mathcal{H}_s^{0,4,-3})_5=\mu^0_{5,9,-3}= [0,0,0,0,1],$$
$$(\mathcal{H}_s^{0,4,3})_5=\mu^0_{5,9,3}= [0,0,0,0,1],$$
$$(\mathcal{H}_s^{0,5,0})_5=\mu^0_{5,10,0}=[0,0,0,0,2],$$
%
$$(\mathcal{H}_s^{0,5,-1})_6=\mu^0_{6,11,-1}=[0,0,0,1,0] + [1,0,0,0,1],$$
$$(\mathcal{H}_s^{0,5,1})_6=\mu^0_{6,11,1}=[0,0,0,1,0] + [1,0,0,0,1],$$
$$(\mathcal{H}_s^{0,5,-3})_6=\mu^0_{6,11,-3}= [0,0,0,1,0],$$
$$(\mathcal{H}_s^{0,5,3})_6=\mu^0_{6,11,3}=[0,0,0,1,0],$$
%
$$(\mathcal{H}_s^{0,6,-1})_7=\mu^0_{7,13,-1}= [0,0,0,0,1],$$
$$(\mathcal{H}_s^{0,6,1})_7=\mu^0_{7,13,1}= [0,0,0,0,1],$$
$$(\mathcal{H}_s^{0,5,-2})_7=\mu^0_{7,12,-2}= [0,1,0,0,0],$$
$$(\mathcal{H}_s^{0,5,2})_7=\mu^0_{7,12,2}= [0,1,0,0,0],$$
$$(\mathcal{H}_s^{0,5,-4})_7=\mu^0_{7,12,-4}= [0,0,0,0,0],$$
$$(\mathcal{H}_s^{0,5,4})_7=\mu^0_{7,12,4}=[0,0,0,0,0],$$
$$(\mathcal{H}_s^{0,5,0})_7=\mu^0_{7,12,0}= [0,0,0,0,0] + [2,0,0,0,0],$$
$$(\mathcal{H}_s^{0,6,-2})_8=\mu^0_{8,14,-2}= [1,0,0,0,0],$$
$$(\mathcal{H}_s^{0,6,2})_8=\mu^0_{8,14,2}= [1,0,0,0,0],$$
$$(\mathcal{H}_s^{0,6,0})_8=\mu^0_{8,14,0}= [1,0,0,0,0],$$
%
$$(\mathcal{H}_s^{0,7,-2})_9=\mu^0_{9,16,-2}=[0,0,0,0,0],$$
$$(\mathcal{H}_s^{0,7,2})_9=\mu^0_{9,16,2}=[0,0,0,0,0],$$
all other groups vanish. The results above are verified by comparing Euler characteristics of
\beq \label{eulerchar:Hs_N=2}
\chi_{p,N,gl}:=\sum_k [(\mathcal{H}^{p,\overline{t},\overline{gl}}_s)_k] (-1)^k=\sum_k (-1)^k [\mu^p_{k,N,\overline{gl}}]
\eeq
with that of
\beq
\chi_{p,N,gl}:=\displaystyle\sum_{i=0}^{N}\displaystyle\sum_{j=-i,j=j+2}^{j} [H^{p,i,j}_t \otimes \Lambda^a(\theta_+)\otimes \Lambda^b(\theta_+)] (-1)^{N-i}
\eeq
where $N$ is defined by Eq.~\ref{N:deg_t_theta}, and
$$a=\frac{N+gl-i-j}{2}, b=\frac{N-gl-i+j}{2}.$$

Generally, a similar formula for higher cohomology group in term of resolutions of modules $H^{p}_t$ (the $p-$th cohomology with respect to the differential $t$)
has the form
$$(\mathcal{H}_s ^{p, \overline{t}, \overline{gl} })_k=\mu^p_{k, \overline{t}+k, \overline{gl} }$$
where $\mu^p_{k, \overline{t}, \overline{gl} }$ stands for $k$-th term in the resolution of $H_t ^{p, \overline{t}, \overline{gl} }.$


\end{document}